\documentclass[epjCONF,columns]{svjour} 
\usepackage{graphics}
\usepackage[varg]{txfonts} 
\usepackage[latin1]{inputenc}
\session-title{NSRT12}
\newcommand{\beq}{\begin{equation}}
\newcommand{\eeq}{\end{equation}}
\newcommand{\bea}{\begin{eqnarray}}
\newcommand{\eea}{\end{eqnarray}}
\newcommand{\eps}{\varepsilon}

\begin{document}
\title{The first quadrupole excitations in spherical nuclei and nuclear pairing}

\author {S.\,V. Tolokonnikov\inst{1}, \inst{2}
\and S. Kamerdzhiev\inst{3} \and S. Krewald\inst{4} \and
E.\,E. Saperstein\inst{1}\fnmsep\thanks{\email{saper@mbslab.kiae.ru}}
\and D. Voitenkov\inst{3} }
\institute{Kurchatov Institute, 123182, Moscow, Russia
\and Moscow Institute of Physics and Technology, 123098 Moscow, Russia
\and  Institute for Physics and Power Engineering, 249033 Obninsk, Russia
\and  Institut f\"ur Kernphysik, Forschungszentrum J\"ulich, D-52425 J\"ulich, Germany }

\abstract {Excitation energies and transition probabilities of the
first $2^+$ excitations in even lead, tin and nickel isotopes are
calculated within the self-consistent Theory of Finite Fermi Systems
based on the Energy Density Functional by Fayans et al. A reasonable
agreement with available experimental data is obtained. The effect
of the density dependence of the effective pairing interaction is
analyzed in detail by comparing results obtained with volume and
surface pairing. The effect is found to be noticeable, especially
for the $2^+$-energies which are systematically higher at 200-300
keV for the volume paring as compared with the surface pairing case,
the latter being in a better agreement with the data.}

\maketitle

\section{Introduction}
Since the famous paper of 1960 by  Belyaev \cite{Bel1}, it became
clear that the low-energy nuclear dynamics is governed mainly with
an interplay of pairing and the quadrupole degree of freedom,
properties of the quadrupole excitations themselves depending
crucially on the pairing. It is clearly seen directly from the
experimental data. Indeed, there is no low-lying collective
$2^+$-state in magic nuclei where pairing is absent. E.g., in the
$^{208}$Pb nucleus the first $2^+$-state has the excitation energy
$\omega_2{=}4.08\;$MeV, but in the neighboring isotope $^{206}$Pb it
decreases to $\omega_2{=}0.80\;$MeV. It is better to say that the
low-lying $2^+$-states and pairing have the common grounds, the
unclosed shell structure where low energy virtual transitions of
positive parity dominate, and pairing is just an approximate way to
take into account  the most important part of these correlations. In
the paper \cite{BE2}, we put attention to effects  of the density
dependence of pairing  on characteristics of $2^+$-states.

Almost 50 years ago, in Ref. \cite{Sap-Tr}, the density-dependent
ansatz for the effective pairing interaction was proposed within the
theory of finite Fermi systems (TFFS) \cite{AB1} with two
(dimensionless) parameters $\gamma_{\rm in}$ and $\gamma_{\rm ex}$
which determine the strength of pairing interaction inside and
outside nucleus correspondingly. Arguments in favor of the version
with the surface dominance, $|\gamma_{\rm in}/\gamma_{\rm ex}|\simeq
10$, were found in \cite{Sap-Tr} and later in \cite{Zver-Sap} on the
base of the detailed analysis of double odd-even mass differences in
spherical nuclei. For brevity, we name it the ``surface pairing'',
in contrast with the ``volume pairing'' model corresponding to
equality $\gamma_{\rm in}{=}\gamma_{\rm ex}$. Very strong support of
the surface version was produced by Fayans et al. \cite{Fay} with
the analysis, within the Energy Density Functional (EDF) method, of
the odd-even staggering effect in charge radii of many
 isotopic chains. The surface pairing version is supported also with
the microscopic theory of pairing \cite{Pankr} starting from a
realistic $NN$ potential. Here, following to \cite{BE2}, we
demonstrate that the systematic, partially new, data on the
excitation energies $\omega_2$ in semi-magic nuclei evidence also in
favor of the surface pairing.

\section{Brief  formalism}
We describe the quadrupole excitations in spherical nuclei within the
self-consistent TFFS \cite{KhS} with the use of the  EDF method of \cite{Fay}.
In this method, the ground
state energy of a nucleus is considered as a functional of normal
and anomalous densities, \beq E_0=\int {\cal E}[\rho_n({\bf
r}),\rho_p({\bf r}),\nu_n({\bf r}),\nu_p({\bf r})] d^3r.\label{E0}
\eeq  This approach is a generalization for superfluid finite systems of the
original Kohn--Sham EDF method \cite{KSh}. The normal part of the EDF ${\cal E}_{\rm norm}$ contains the
finite-range central, spin-orbit and effective tensor nuclear terms and Coulomb
interaction term for protons. We use the DF3-a functional \cite{Tol-Sap} which
is a slight modification of the DF3 functional of \cite{Fay} concerning the
spin-orbit and effective tensor terms.

Within the TFFS, the response of a nucleus to the external quadrupole
field $V_0\exp{(i\omega t)}$ can be found in terms of the effective field.
In systems with pairing correlations, equation  for the effective
field can be written in a compact form as \beq {\hat
V}(\omega)={\hat V}_0(\omega)+{\hat {\cal F}}  {\hat A}(\omega) {\hat V}(\omega), \label{Vef_s}
\eeq where all the terms  are  matrices. In the
standard TFFS notation \cite{AB1}, we have:
\beq {\hat V}=\left(\begin{array}{c}V
\\d_1\\d_2\end{array}\right)\,,\quad{\hat
V}_0=\left(\begin{array}{c}V_0
\\0\\0\end{array}\right)\,,
\label{Vs} \eeq

\beq {\hat {\cal F}}=\left(\begin{array}{ccc}
{\cal F} &{\cal F}^{\omega \xi}&{\cal F}^{\omega \xi}\\
{\cal F}^{\xi \omega }&{\cal F}^\xi  &{\cal F}^{\xi \omega }\\
{\cal F}^{\xi \omega }&{\cal F}^{\xi \omega }& {\cal F}^\xi \end{array}\right), \label{Fs} \eeq

\beq {\hat A}(\omega)=\left(\begin{array}{ccc} {\cal L}(\omega) &{\cal M}_1(\omega)
&{\cal M}_2(\omega)\\
 {\cal O}(\omega)&-{\cal N}_1(\omega) &{\cal N}_2(\omega)\\{\cal O}(-\omega)&-{\cal N}_1(-\omega) &
 {\cal N}_2(-\omega)
\end{array}\right)\,,
\label{As} \eeq where ${\cal L},\; {\cal M}_1$, and so on stand
for integrals over $\eps$ of the products of different
combinations of two Green functions $G(\eps)$ and two Gor'kov
functios $F^{(1)}(\eps)$ and $F^{(2)}(\eps)$. They can be found in
\cite{AB1} and we write down here only the first of them which is
of the main importance for us, \beq {\cal
L}=\int\frac{d\varepsilon}{2\pi
i}\left[G(\varepsilon)G(\varepsilon+\omega)-F^{(1)}(\varepsilon)F^{(2)}(\varepsilon+\omega)
\right]. \label{Ls} \eeq

Isotopic indices in Eqs. (\ref{Vs}-\ref{Ls}) are omitted for
brevity. In Eq. (\ref{Fs}), ${\cal F}$ is the usual Landau--Migdal
amplitude, \beq {\cal F}=\frac {\delta^2 {\cal E}}{\delta \rho^2},
\label{LM} \eeq ${\cal F^{\xi}}$ is the effective pairing
interaction, \beq {\cal F^{\xi}}=\frac {\delta^2 {\cal E}}{\delta
\nu^2}, \label{LM} \eeq whereas the amplitudes ${\cal F}^{\omega
\xi}={\cal F}^{\xi \omega}$ stand for the mixed second derivatives,
\beq {\cal F}^{\omega \xi}=\frac {\delta^2 {\cal E}}{\delta \rho
\delta \nu}. \label{LMxi} \eeq  In the case of volume pairing, we
have ${\cal F}^{\omega \xi}=0$. The explicit form of the above
equations is written down for the case of the electric ($t$-even)
symmetry we deal with. A static moment of an odd nucleus can be
found in terms of the diagonal matrix element $ \langle\lambda_0|
V(\omega =0)|\lambda_0\rangle$ of the effective field over the state
$\lambda_0$ of the odd nucleon.

The effective field operator ${\hat V}(\omega)$ has a pole in the excitation
energy $\omega_s$ of the state $|s\rangle$ under consideration,
\begin{equation}\label{pole}
    {\hat V}(\omega)=\frac {\left({\hat V_0}{\hat A(\omega_s)} {\hat g}_{0s}
     \right){\hat g}_{0s}}{\omega-\omega_s}+{\hat V}_R(\omega).
\end{equation}
The quantity ${\hat g}_{0s}$ has the meaning of the corresponding
excitation amplitude. It obeys  the homogeneous counterpart of Eq.
(\ref{Vef_s}) and is normalized as follows \cite{AB1},
\begin{equation}\label{norm}
\left({\hat g}_{0s}^+ \frac {d {\hat A}}{d\omega}{\hat g}_{0s} \right)_{\omega=\omega_s}=-1,
\end{equation}
with obvious notation.

It should be noted that the above TFFS equations are essentially
equivalent to QRPA those. Important difference occur if one goes
beyond the QRPA \cite{Voit}. For the pairing, we use the diagonal
approximation, $\Delta_{\lambda\lambda'}{=}
\Delta_{\lambda}\delta_{\lambda\lambda'}$, and ``developed pairing''
formalism which neglects effects of particle non-conservation.
According to \cite{part-numb}, it leads to errors $\simeq 0.1\;$MeV
in values of $\Delta_{\lambda}$.

\section{Surface {\it versus} volume pairing for semi-magic nuclei}

\begin{figure}
\resizebox{1.00\columnwidth}{!} {\includegraphics {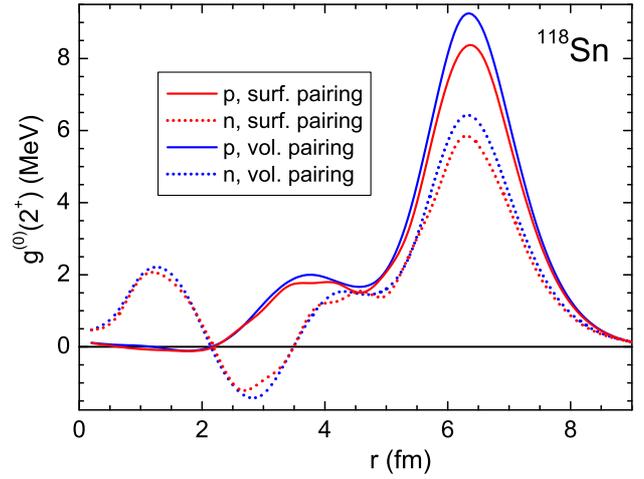}}
\caption{The proton and neutron normal transition amplitudes
$g^{(0)}$ in $^{118}$Sn nucleus.} \label{fig:gnorm}
\end{figure}

\begin{figure}
\resizebox{1.00\columnwidth}{!} {\includegraphics{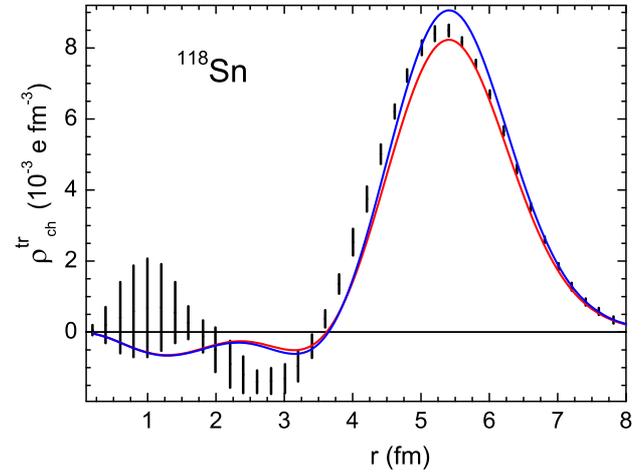}}
\vspace{1mm} \caption{The charge transition densities $\rho_{\rm
ch}^{{\rm tr}(0)}$ in $^{118}$Sn  nucleus. Red line corresponds to
 surface pairing, blue one, to volume pairing. Experimental
data are taken from \cite{rhotr}.} \label{fig:rhoch}
\end{figure}

The QRPA-like equations of the self-consistent TFFS were solved in the coordinate
representation, with exact consideration of continuum.
   We limit ourselves with semi-magic nuclei where the phonon coupling corrections
to QRPA are less than in nuclei with both non-magic subsystems. As a benchmark nucleus,
we take  $^{118}$Sn which is in the middle of the tin chain.  Let us begin from the  normal
components. As it is seen from Fig. \ref{fig:gnorm}, the normal transition amplitudes $g^{(0)}(r)$
found for these two kinds of pairing are rather close to each other. The same is true for the
normal components of transition densities that is illustrated for the charge transition density
$\rho_{\rm ch}^{{\rm tr}(0)}(r)$  which is compared in Fig. \ref{fig:rhoch} with the experimental
one found in Ref. \cite{rhotr} with the model-independent analysis of the high-precision data on the elastic
electron scattering. It is worth mentioning that these two figures confirm the interpretation by Khodel
\cite{Khod-cap,KhS} of the $2^+_1$-excitations as quantum capillary waves belonging to the Goldstone branch
of surface vibrations.

\begin{figure}
\resizebox{1.00\columnwidth}{!} {\includegraphics {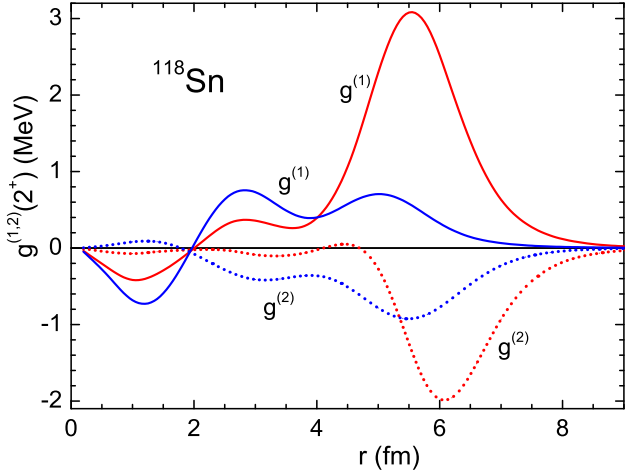}}
\caption{The neutron anomalous transition amplitudes $g^{(1,2)}$ in
$^{118}$Sn nucleus. Red solid and dotted  lines correspond to
 surface pairing, blue ones, to volume pairing.}
\label{fig:abnorm}
\end{figure}

Another situation appears for the anomalous transition amplitudes
$g^{(1,2)}$, see Fig. \ref{fig:abnorm}. We see that now these
components in the case of the surface pairing are approximately 3
times stronger than for volume pairing. Unfortunately, there is no
direct way to observe the anomalous transition amplitudes
$g^{(1,2)}$, but there is an indirect one. When comparing Fig.
\ref{fig:gnorm} and \ref{fig:abnorm}, one should take into account
that the amplitudes $g^{(0)}$ and $g^{(1,2)}$ have the same
normalization, therefore their absolute values may be compared
directly. Thus, we see that in the nucleus under consideration the
anomalous components are comparatively small, approximately 3 times
smaller than the normal one even for the surface pairing case.
Therefore we  can consider the anomalous components  as a small
perturbation and present the eigen-energy of the $L$-phonon as \beq
\omega_L{=}\omega_L^{(0)}{+}\delta\omega_L^{(1)}{+}\delta\omega_L^{(2)},\eeq
where the corrections due to the anomalous components are \beq
\delta\omega_L^{(1,2)}\propto -(g^{(1,2)})^2/\omega_L.
\label{del-om}\eeq Here $g^{(1,2)}$ denote  average values of the
corresponding matrix elements. Therefore, for the surface pairing
this (negative) correction should be larger. It turns out that way
in actual calculations. For  $B(E2)$ values, there is an
interference between the normal amplitude  $g^{(0)}$ and the
anomalous ones, $g^{(1,2)}$, therefore dependence of the result on
the type of pairing is not so definite.

\begin{figure}
\resizebox{1.00\columnwidth}{!} {\includegraphics {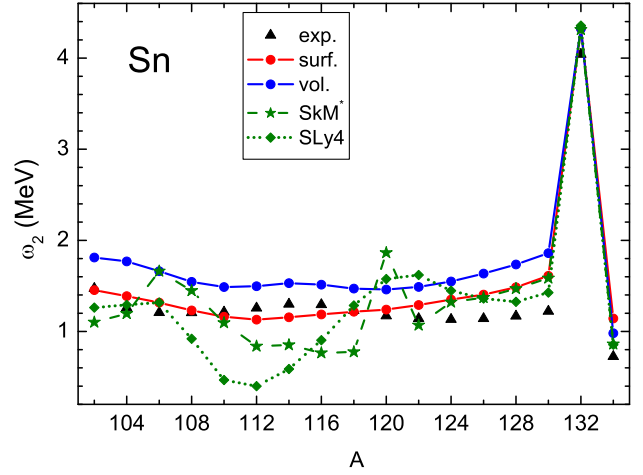}} \caption{
Excitation energies $\omega(2^+_1)$  for tin isotopes. Experimental
data are taken from \cite{Dat}.} \label{fig:Sn-E2}
\end{figure}

Calculation results for $\omega_2$  values in tin isotopes are
displayed in Fig. \ref{fig:Sn-E2}. We see that for the surface
pairing $\omega_2$ values are approximately by 0.3 MeV less than for
the volume one, in a qualitative accordance with Eq. (\ref{del-om}).
The surface version turns out to be closer to the experiment. The
rms deviation of the theory from the experiment  is ${\Delta
\omega}_{\rm rms}{=}0.16$  MeV for the surface pairing and ${\Delta
\omega}_{\rm rms}{=}0.37$  MeV for the volume pairing. When
calculating the rms values of ${\Delta \omega}_{\rm rms}$, we
excluded the double magic nuclei and their neighbors, $^{102}$Sn and
$^{130-134}$Sn. Magic nuclei are excluded on the reason of absence
of pairing and hence the effect under consideration. As to their
neighbors, the reason is deficiencies of  the ``developed pairing''
approximation we use for such nuclei. First, the particle number
conservation only in average inherent to this approximation works
for them  rather poorly, as it follows from the analysis in Ref.
\cite{part-numb}. Second, it is true for the equality of the
operators $\Delta^+$ and $\Delta^-$, see Ref. \cite{AB1}, which is
supposed in this approximate scheme. Indeed, e.g. for an isotope
with the neutron number $N=N_{\rm mag}-2$, the operator $\Delta^+$
connects the nucleus under consideration with the one in which the
neutron pairing is absent.

\begin{figure*}
\resizebox{2.00\columnwidth}{!} {\includegraphics {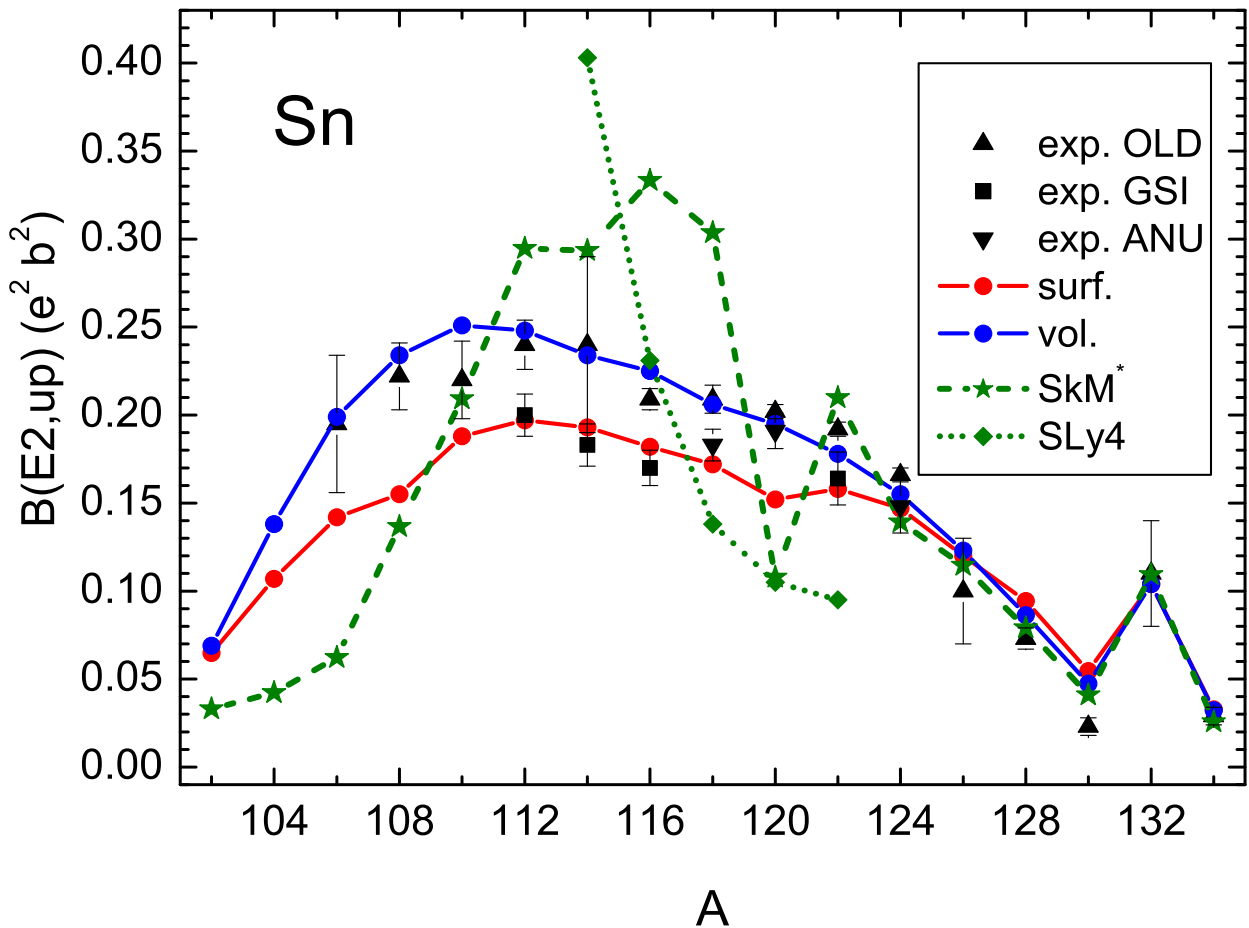}} \caption{
$B(E2,{\rm up})$ values for tin isotopes.
 The ``old'' experimental
data are taken for $^{114-124}$Sn from \cite{Dat}, for
$^{126-134}$Sn  from \cite{newbe2}, and for $^{106-112}$Sn from
\cite{be2_110,be2_106-112,be2_106-108}. The new values marked
``GSI'' and ``ANU'' are taken from \cite{GSI-be2}. The first values
are measured at the at the UNILAC accelerator of the Gesellschaft
f$\ddot{u}$r
 Schwerionenforschung (GSI). The second ones are found
also in \cite{GSI-be2} by analyzing data obtained in the magnetic
moment measurement \cite{ANU-be2} at the Australian National
University (ANU).} \label{fig:Sn-BE2}
\end{figure*}

\newpage
The $B(E2)$ values in the tin chain are shown in   Fig.
\ref{fig:Sn-BE2}. In this case, the situation is not so definite
because of experimental errors. It is worth mentioning that in this
case the volume version looked a bit better when the old data were
used \cite{BE2}, but new data of \cite{GSI-be2} agree better with
the surface version. A comparison is made with the
Skyrme--Hartree--Fock--Bogolyubov (SHFB) theory predictions of Ref.
\cite{Bertsch1} with the SkM* and SLy4 Skyrme force. Among these two
SHFB results, the SkM* ones are significantly better.  Our results
for the excitation energy, especially for the surface pairing, are
much better than both the SHFB ones. The same is true for the
$B(E2)$ values. A general deficiency of the two SHFB theory
predictions is their non-regular $A$-dependence whereas the data
depend on $A$ rather smoothly.

\begin{figure}
\resizebox{1.00\columnwidth}{!} {\includegraphics {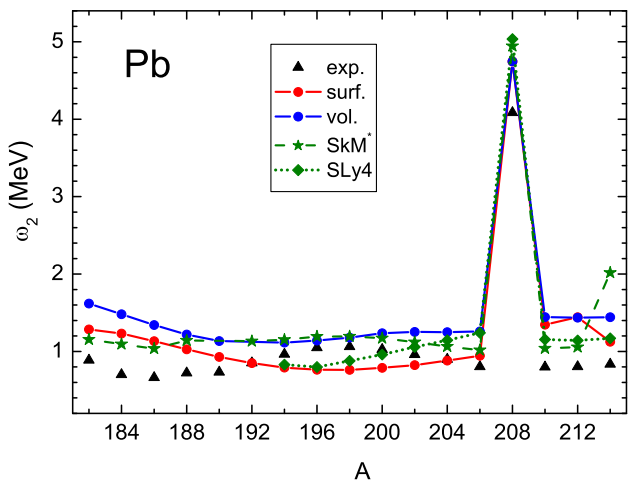}} \caption{
Excitation energies $\omega(2^+_1)$  for lead isotopes.}
\label{fig:Pb-E2}
\end{figure}

\begin{figure}
\resizebox{1.00\columnwidth}{!} {\includegraphics {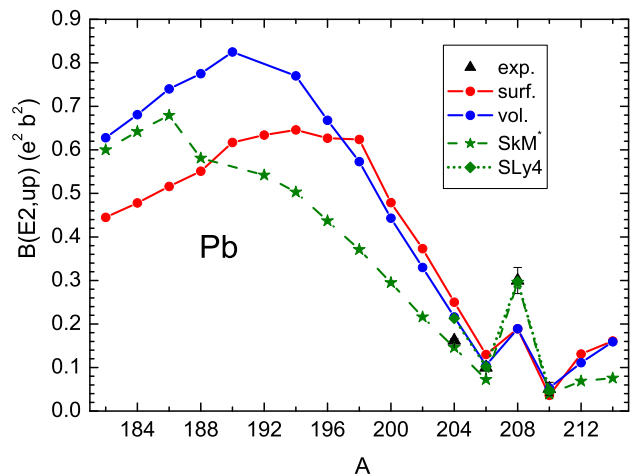}}
\caption{$B(E2,{\rm up})$ values for lead isotopes.}
\label{fig:Pb-BE2}
\end{figure}

For the lead isotopes, the situation is qualitatively similar, see
Figs. \ref{fig:Pb-E2}, \ref{fig:Pb-BE2}. Again, $\omega_2$ values
for the volume pairing are at 0.3 MeV higher of those for the
surface case. Now we have ${\Delta \omega}_{\rm rms}{=}0.33$ MeV for
surface pairing and ${\Delta \omega}_{\rm rms}{=}0.47$ MeV for
volume pairing. In accordance with the above arguments, the
calculation of the ${\Delta \omega}_{\rm rms}$ values is carried out
with exclusion of  $^{206-210}$Pb nuclei. We see that here accuracy
of our calculations is worse than for the tin chain but again the
surface pairing model predictions look better. Here the SHFB results
for both the versions of Skyrme force look much better than for the
tin chain and their accuracy is approximately the same as ours for
the surface pairing case.  The $B(E2)$ values are known only for
four lead isotopes, $^{204-210}$Pb, the SHFB accuracy is a bit
higher. Note that  the $^{204}$Pb nucleus is the only among them
which is described within our QRPA-like scheme adequately.

\begin{figure}
\resizebox{1.00\columnwidth}{!} {\includegraphics {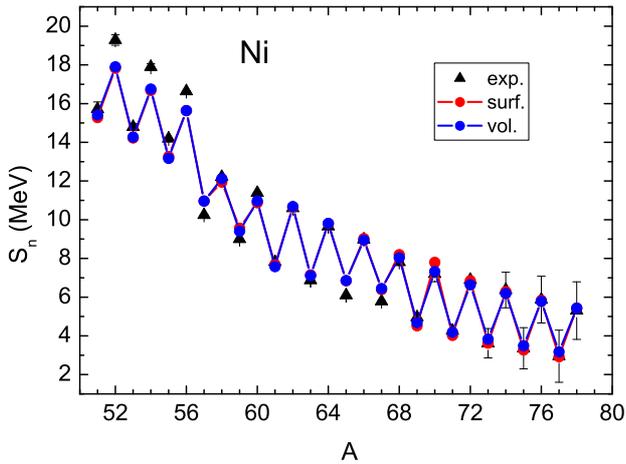}} \caption{
Neutron separation energies $S_n$ for Ni isotopes.}
\label{fig:Ni-Sn}
\end{figure}

\begin{figure}
\resizebox{1.00\columnwidth}{!} {\includegraphics {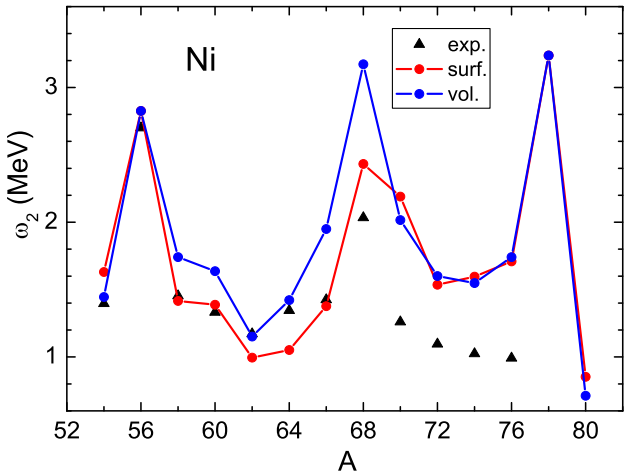}} \caption{
Excitation energies $\omega(2^+_1)$ for Ni isotopes.}
\label{fig:Ni-E2}
\end{figure}

\begin{figure}
\resizebox{1.00\columnwidth}{!} {\includegraphics {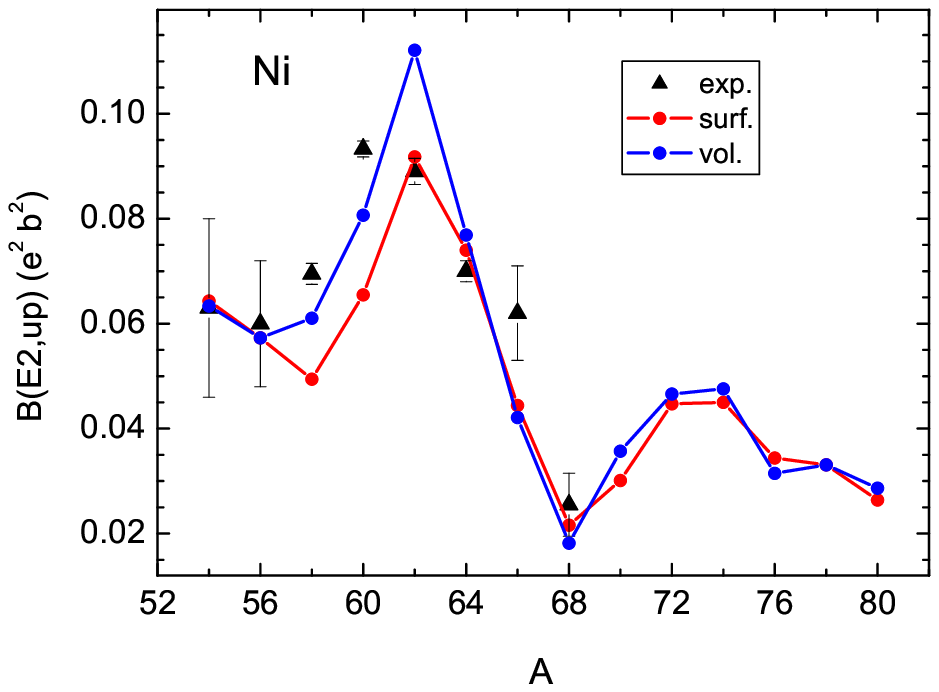}}
\caption{$B(E2,{\rm up})$ values for Ni isotopes} \label{fig:Ni-BE2}
\end{figure}

Let us go to the nickel isotopes.
  Fig. \ref{fig:Ni-Sn} demonstrates that the two versions
of the pairing force we compare are practically equivalent to each
other from the point of view of reproducing  the data on the neutron
separation energy $S_n$ for the nickel chain. Analogous
demonstration for the tin and lead isotopes can be found in
\cite{BE2}. Calculation results for excitations energies are
displayed in Fig. \ref{fig:Ni-E2} and for $B(E2)$ values, in Fig.
\ref{fig:Ni-BE2}. The chain contains two double-magic nuclei,
$^{56}$Ni, $N=28$,  and $^{78}$Ni, $N=50$. In addition, the nucleus
$^{68}$Ni with the neutron number $N=40$ also can be interpreted as
a quasi-magic one. Indeed, the excitation energy of the $2^+_1$
state in this nucleus is significantly higher than in neighboring
even nickel isotopes that is a typical feature of magic nuclei.
Although the effect is not so strong as for $^{56}$Ni, it is also
rather pronounced. For a complete magicity, the distance between
occupied single-particle levels and the next free one  should be
greater than the double average gap value $2\bar{\Delta}$. In this
case, the gap equation has only the trivial solution. In the case
under discussion, the distance between the occupied  $2p-1f$-shell
and the next level $1g_{9/2}$ is of the order of $2\bar{\Delta}$ and
the gap equation has a non-trivial solution. However, some features
of magic nuclei occur in this nucleus but in a smoothed form. Thus,
just as for the tin and lead chain, for analyzing the ``surface {\it
versus} volume'' effect under discussion, it is worth to exclude the
neighboring to $^{56}$Ni and $^{78}$Ni nuclei. In addition, the
isotopes $^{66-70}$Ni should be also considered with some caution.

\begin{figure}
\resizebox{1.00\columnwidth}{!} {\includegraphics {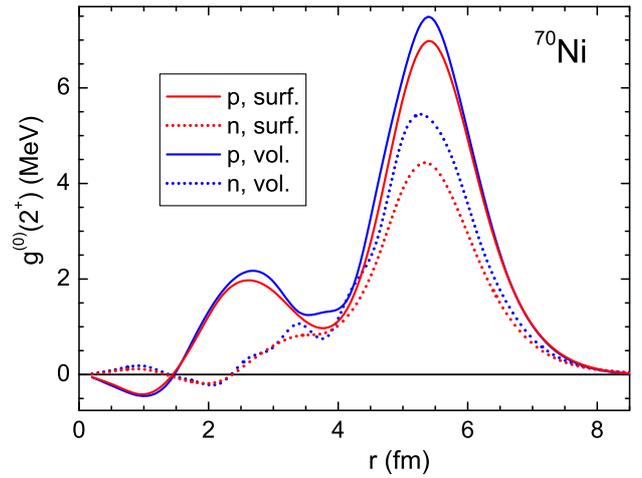}}
\caption{The proton and neutron normal transition amplitude
$g^{(0)}$ in the $^{70}$Ni isotope.} \label{fig:Ni70-g0}
\end{figure}

\begin{figure}
\resizebox{1.00\columnwidth}{!} {\includegraphics {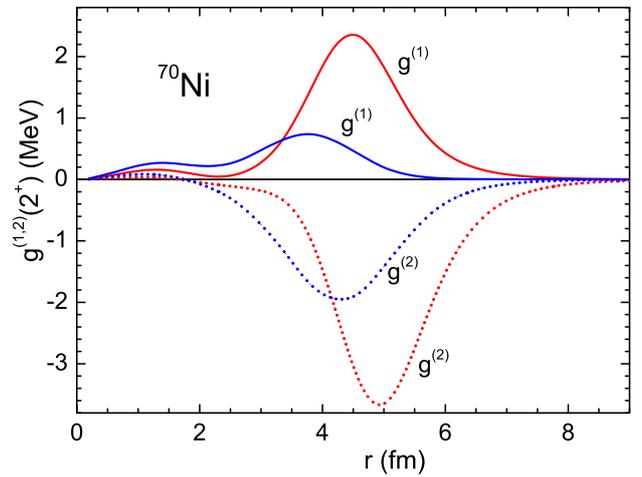}}
\caption{The neutron anomalous  transition amplitudes $g^{(1)}$ and
$g^{(2)}$ in the $^{70}$Ni isotope. Red solid and dotted lines
correspond to surface pairing, blue ones, to volume pairing.}
\label{fig:Ni70-g12}
\end{figure}

We see that qualitatively the pattern is similar to that in the tin
and lead chains, however the effect for $\omega_2$ value is now not
so regular. Indeed, for the isotopes $^{72,74}$Ni it is practically
zero and for the isotope $^{70}$Ni it is of the opposite sign,
although rather small.  Although the latter could be interpreted as
the effect of closeness to the quasi-magic nucleus $^{68}$Ni, let us
try to understand the reason of such contradiction to the formula
(12) within the scheme we use. For this aim, we displayed for this
nucleus  the normal and anomalous transition amplitudes  in Fig.
\ref{fig:Ni70-g0} and Fig. \ref{fig:Ni70-g12} correspondingly. We
see that, just as for the $^{118}$Sn nucleus, the normal amplitudes
for the two models of pairing under consideration are very close to
each other. In addition, again the anomalous amplitudes for the
surface pairing case are much stronger than those for the volume
pairing. However, in this nucleus the absolute values of the
anomalous amplitudes are rather close to those of the normal
amplitudes. Therefore, the consideration of the anomalous amplitudes
as a small perturbation, which is the base for receiving the
estimation (12), is not valid now.

Agreement with the data on $\omega_2$ values for the surface pairing
case is almost perfect for nuclei lighter than $^{68}$Ni, however it
becomes worse for heavier isotopes. The most strong disagreement for
$^{70}$Ni and $^{76}$Ni nuclei again can be interpreted as  the
effect of closeness to  magic nuclei (a quasi-magic in the first
case). One more reason  exists for poorer agreement for heavy nickel
isotopes. For them, neutrons are filling up an almost isolated level
$1g_{9/2}$. Indeed, as it was already discussed, the distance to the
filled levels of negative parity is about $2\bar{\Delta}$ whereas
that to neighboring levels of positive parity is significantly
greater than $2\bar{\Delta}$. In such a situation, predictions for
the positive parity states depend significantly on details of the
calculation scheme. The experimental values of transition
probabilities are known only for isotopes till $^{68}$Ni, here
agreement is quite reasonable.

\section{Conclusion}
Excitation energies,  transition probabilities and other
characteristics of the first $2^+$ excitations in three chains of
semi-magic isotopes, the lead, tin and nickel ones, are calculated
within the self-consistent TFFS on the base of the EDF  by Fayans et
al. \cite{Fay}. The DF3-a functional \cite{Tol-Sap} is used which
differs from the original one DF3 with the spin-orbit and effective
tensor terms only.  A reasonable agreement with available
experimental data is obtained. Predictions for other isotopes are
made.

The effect of the density dependence of the effective pairing
interaction is analyzed  by comparing predictions of two models of
the effective pairing interaction, the surface pairing and the
volume one. The effect is found to be noticeable, especially for the
$2^+$-energies which are, with the only exception, systematically
higher at 200-300 keV for the volume paring as compared with the
surface pairing case. Thereby predictions of the surface pairing
model agree with the data better than those of the volume model.
Thus, the analysis of the excitation energies of the first $2^+$-
states in semi-magic nuclei gives a new evidence in favor of the
surface pairing. As to $B(E2)$ values in tin isotopes, the volume
pairing model looked better in comparison with old experimental
data, but the new data of \cite{GSI-be2} agree better with the
surface model.

\section{Acknowledgment}
We thank N. Pietralla  for valuable discussion and information about
new data on $B(E2)$ values in tin isotopes. The work was partly
supported by the DFG and RFBR Grants Nos.436RUS113/994/0-1 and
09-02-91352NNIO-a, by the Grants NSh-7235.2010.2  and 2.1.1/4540 of
the Russian Ministry for Science and Education, and by the RFBR
grants 11-02-00467-a and 12-02-00955-a.

\end{document}